# Liquid-Gated High Mobility and Quantum Oscillation of the Two-Dimensional Electron Gas at an Oxide Interface

Shengwei Zeng,[†,‡] Weiming Lü,[†,■] Zhen Huang,[†] Zhiqi Liu,[†,▲] Kun Han,[†,‡] Kalon Gopinadhan,[†] Changjian Li,[†,§] Rui Guo[†,#], Wenxiong Zhou,[†,‡] Haijiao Harsan Ma,[†,‡] Linke Jian,[†] Thirumalai Venkatesan,[†,‡,§,#,□,*] Ariando[†,‡,§,*]

[†]NUSNNI-NanoCore, National University of Singapore, Singapore 117411, Singapore
[‡]Department of Physics, National University of Singapore, Singapore 117542, Singapore
[§]National University of Singapore Graduate School for Integrative Sciences and Engineering (NGS), 28 Medical Drive, Singapore 117456, Singapore
[#]Department of Materials Science and Engineering, National University of Singapore, Singapore 117575, Singapore
[□]Department of Electrical and Computer Engineering, National University of Singapore, Singapore 117576, Singapore
[■]Present address: Department of Science, Condense Matter Science and Technology Institute, Harbin Institute of Technology, Harbin 150001, China
[▲]Present address: Oak Ridge National Laboratory, Oak Ridge, Tennessee 37831, United States

*To whom correspondence should be addressed.
E-mail: venky@nus.edu.sg; ariando@nus.edu.sg

## ABSTRACT

Electric field effect in electronic double layer transistor (EDLT) configuration with ionic liquids as the dielectric materials is a powerful means of exploring various properties in different materials. Here we demonstrate the modulation of electrical transport properties and extremely high mobility of two-dimensional electron gas at $LaAlO_3/SrTiO_3$ (LAO/STO) interface through ionic liquid-assisted electric field effect. By changing the gate voltages, the depletion of charge carrier and the resultant enhancement of electron mobility up to 19380 $cm^2/Vs$ are realized, leading to quantum oscillations of the conductivity at the LAO/STO interface. The present results suggest that high-mobility oxide interfaces which exhibit quantum phenomena could be obtained by ionic liquid-assisted field effect.

Electric field effect doping has been a flexible and powerful technique to tune the carrier density and the resultant transport properties of a material.[1] More recently, the field effect in EDLT configuration with ionic liquids (ILs) and polymer electrolytes as gate dielectrics has attracted growing interest. In an EDLT, IL is applied directly on the surface of interest and polarized by gate voltage, inducing a two-dimensional (2D) layer of charges on the sample surface with high carrier density of up to $10^{15}/cm^2$.[2] Owing to the capability of EDLTs in accumulating large amount of carriers, electric field-induced phase transitions have been demonstrated in various materials.[2-7] In the push for practical application of ILs-assisted field-effect transistors, it is desirable to obtain high carrier mobility at the 2D layer. However, since the ILs are applied on the sample surface, disorder is induced due to surface-degrading electrochemical reactions,[7-9] which hinders the improvement of the mobility. For example, the reported mobility is ~1000 $cm^2/Vs$ at IL/SrTiO$_3$,[3] ~250 $cm^2/Vs$ at IL/ZnO,[2] ~7000 $cm^2/Vs$ at IL/KTaO$_3$ interfaces below 10 K,[4] and ~330 $cm^2/Vs$ at interfaces between IL and 2D materials at 77 K.[10-13] Therefore, protection of surface would help to improve the mobility of the induced 2D charged layer.[14] For the 2D electron gas (2DEG) at the interface between two insulators formed by depositing thin film on substrate, it is expected that the mobility can be improved by IL gating since the deposited capping layer can be used as the protection layer. Here, we demonstrate a high-mobility 2DEG at LAO/STO interface as a result of IL gating.

Among oxide interfaces, LAO/STO interface is of particular interest since it exhibited fascinating properties such as conductivity,[15] superconductivity,[16] magnetism,[17] electronic phase separation,[18] 2D quantum oscillations,[19-22] and controversial origin of conductivity.[23] These phenomena provide new insights for understanding the nature of electronic and magnetic properties at strongly correlated oxide interfaces. Enhancements of mobility at LAO/STO interfaces and STO-based heterostructures have been explored through defect engineering,[24] surface control,[25] dimensionality control,[26] charge-transfer-induced modulation doping,[27] and growth strategies.[19,28-31] Electric field effect could modulate charge carriers and be flexibly implemented on devices, it is therefore vital to explore its tunability of the LAO/STO interfaces. Electric field induced insulator-to-metal and superconductor-to-insulator transitions have been demonstrated using STO,[32-34] LAO,[35-39] ferroelectric $Pb(Zr_{0.2}Ti_{0.8})O_3$,[40,41] $Si_3N_4$ [42] and IL[43] as the dielectric materials. However, to the best of our knowledge, there is no report on significant enhancement of mobility at the LAO/STO interface by means of electric field effect; only a slight enhancement (from ~1000 to ~2000 $cm^2/Vs$) or even reduction in mobility was observed.[34, 35, 38] In this paper, we demonstrate the modulation of electrical transport properties of LAO/STO interface by electric field effect in EDLT configuration. Through changing the gate voltage, the carrier density at the interface is depleted, and therefore, the mobility is significantly enhanced up to 19380 $cm^2/Vs$ at low temperature. Due to enhancement of mobility, clear quantum oscillations of the conductance is observed.

**RESULTS AND DISCUSSION**

Figure 1a shows an optical micrograph of a typical pattern on STO for field effect measurement. The width of the Hall bar is 50 μm (some of devices have width of 200 μm) and length is 500 μm, the distance between two voltage probes is 160 μm. A six-probe configuration of the Hall bar allows for the measurement of both longitudinal and Hall resistance. Figure 1b shows an atomic force microscopy (AFM) image of 10-uc LAO grown on the patterned STO substrate. Clear atomic terraces are observed, indicating the smooth surface of patterned substrate and high-quality thin films. Figure 1c-d show source-drain current ($I_D$), leakage current ($I_G$) and resistance as a function of gate voltage ($V_G$). It can be seen that $I_G$ is within 1 nA and independent of $V_G$. This negligibly small $I_G$ indicates a good operation of the device. $I_D$ is significantly higher than $I_G$ at high $V_G$, and decreases with decreasing $V_G$ down to the value of $I_G$ at $V_G \approx -0.6$ V. Resistance increases with decreasing $V_G$ up to the measurement limit at $V_G \approx -0.6$ V. The $V_G$ needed to obtain the insulating state is significantly lower than that in back-gating device (which is on the order of 100 V).[32-34] Scanning the $V_G$ between -0.8 and 2 V, reversible metal-insulator transition with an ON/OFF ratio of ~$10^4$ is observed and the transition is non-hysteretic, which can be seen from the fact that the curves of three scan cycles are significantly similar.

To fully demonstrate the validity of mobility enhancement by IL gating, four samples (samples A, B, C and D) with different initial low-temperature mobilities were synthesized and characterized (see Methods). Figure 2a shows the sheet resistance *versus* temperature ($R_s$-$T$) curves at various $V_G$ for sample A. The $R_s$-$T$ curves for samples B, C and D are shown in Supporting Information Fig. S2. Before the $R_s$-$T$ curves measurements, the $V_G$ was applied at 230 K and kept for 10 minutes for charging. Resistance measurement was made as the samples were cooled down while keeping the $V_G$ constant. In order to change $V_G$, the sample was heated to 230 K after each $R_s$-$T$ curve measurement and a new $V_G$ was applied. At $-0.4 \leq V_G \leq 3$ V, the sample exhibits metallic behaviour, showing a decrease in $R_s$ as $T$ is reduced. At high $T$, as the $V_G$ is increased, $R_s$ decreases monotonously (Fig. 2a). This indicates that increasing $V_G$ causes enhancement of carrier density at the interface. However, at low $T$, $R_s$ does not show monotonous change with increasing $V_G$. $R_s$ decreases as $V_G$ is increased from -0.4 to 0 V while increases as $V_G$ is increased from 0 to 3 V (Inset of Fig. 2a). At $V_G \geq 1.5$ V, $R_s$ shows a slight upturn at $T$ below ~5 K, suggesting localization at high $V_G$. However, at $-0.4 \leq V_G \leq 1$ V the sample shows completely metallic behaviour down to 2 K, suggesting that charge carriers are less scattered at this $V_G$ regime.

The carrier density $n_s$ as a function of $T$ at various $V_G$ for sample A is shown in Fig. 2b. At high $T$ regime, $n_s$ increases with increasing $V_G$. However, at low $T$ regime from 20 to 2 K, $n_s$ increases with increasing $V_G$ from -0.4 to 1 V and saturates at ~$3 \times 10^{13}$ cm$^{-2}$ for $V_G \geq 1.5$ V. The saturation of $n_s$ at 2 K can be clearly seen in Fig. 3a. For the dependence of $n_s$ on $T$, behaviour of $n_s$ is different between two $V_G$ regimes. For $V_G \leq 1$ V, $n_s$ is nearly $T$ independent at each $V_G$. For $V_G \geq 1.5$ V, $n_s$ generally decreases with decreasing $T$, suggesting the carrier freeze-out effect. In contrast to $n_s$, Hall mobility $\mu_H$ increases with decreasing $T$ at all $V_G$, as shown in Fig. 2c. For example, at $V_G = -0.4$ V, $\mu_H$ increases from ~24 cm$^2$/Vs at 180 K to ~6600 cm$^2$/Vs at 2 K. For the dependence of $\mu_H$ on $V_G$, at high $T$ of 180 K, $\mu_H$ exhibits little $V_G$ dependence, changing from 20 cm$^2$/Vs at $V_G = 2$ V to 24 cm$^2$/Vs at $V_G = -0.4$ V. However, at low $T$, a significant change of $\mu_H$ is observed. For example, at $T = 2$ K, $\mu_H$ increases from ~650 cm$^2$/Vs at $V_G = 2$ V to ~6600 cm$^2$/Vs at $V_G = -0.4$ V (see also Fig. 3a).

The $R_s$-$T$, $n_s$-$T$ and $\mu_H$-$T$ curves for the sample A without IL on top of LAO are also plotted in Fig. 2. One can see that the curves for the sample without IL and for the sample with IL but without application of $V_G$ ($V_G = 0$ V) do not coincide with each other. The sample with $V_G = 0$ V show higher $R_s$ at high $T$ but lower $R_s$ at low $T$, compared with that for sample without IL. Moreover, at $V_G = 0$ V the $n_s$ is lower and $\mu_H$ is higher than that without IL. These suggest that there is a coupling between the interface and surface which was covered by IL even at $V_G = 0$ V, causing depletion of $n_s$ at the interface. The interface-surface coupling and the resultant change of electrical transport properties of LAO/STO heterostructures have also been observed by capping polar chemical solvents such as acetone, ethanol and water on top of LAO surface.[25, 44, 45] Enhancements of $n_s$ and reduction of $\mu_H$ were obtained by covering the polar solvents.[44] However, our results show the opposite behaviour with reduction of $n_s$ and enhancement of $\mu_H$. This may be due to the different properties between polar solvents and IL.

Figure 3 shows the $n_s$, $\mu_H$ as a function of $V_G$ and $\mu_H$ as a function of $n_s$ at low $T$ for four samples with different initial $n_s$ and $\mu_H$. One can see that for all samples, significant depletion of $n_s$ and enhancement of $\mu_H$ are obtained by increasing negative $V_G$. Considering that $\mu_H = 1/(e \cdot n_s \cdot R_s)$, where $e$ is the elementary charge, and that $R_s$ does not change much with $V_G$ at low $T$ (Fig. 2a and Supporting Information Fig. S2), we are able to attribute enhancement of $\mu_H$ to the depletion of $n_s$. This can be clearly seen in Fig. 3c, showing that $\mu_H$ increases with decreasing $n_s$ over a broad $n_s$ range. The highest mobilities obtained here are 6600 cm$^2$/Vs at $5\times10^{12}$ cm$^{-2}$, 11330 cm$^2$/Vs at $1.2\times10^{12}$ cm$^{-2}$, 13320 cm$^2$/Vs at $2.1\times10^{12}$ cm$^{-2}$ and 19380 cm$^2$/Vs at $3.3\times10^{12}$ cm$^{-2}$ for samples A, B, C and D, respectively, which are significantly higher than those before IL gating, 1110, 2900, 3500 and 9300 cm$^2$/Vs for samples A, B, C and D, respectively. The data without IL are also plotted in Fig. 3. As discussed above, due to the interface-surface coupling, depletion of $n_s$ and enhancement of $\mu_H$ after covering IL are observed. Compared with the ones without IL, after covering IL ($V_G = 0$ V), $\mu_H$ increases from 1110 to 3400 cm$^2$/Vs, from 2900 to 7060 cm$^2$/Vs and from 3500 to 7360 cm$^2$/Vs for samples A, B and C, respectively. For sample D, after covering IL, it shows highly insulating at low $T$ and the resistance is out of the measurement limit at $V_G = 0$ V. This may be because sample D has lower initial $n_s$ (6.6×10$^{12}$ cm$^{-2}$), leading to a much lower $n_s$ as the IL is applied on the surface of the sample, which could not support the metallic behaviour. Upon increasing the $V_G$ and therefore increasing $n_s$, sample D shows metallic behaviour and highest mobility of 19380 cm$^2$/Vs at $T = 2.8$ K for $V_G = 0.075$ V.

It has been shown that IL gating causes degradation of the sample surface due to the electrochemical reaction.[7-9, 46] However, the AFM images of the sample surfaces before and after gating experiments show that the surfaces are significantly similar (Supporting Information Fig. S3). Moreover, the IL-gated metal-insulator transition is reversible and non-hysteretic, and the leakage current is significantly small (Fig. 1c-d). These results suggest that the interfaces do not degrade after IL gating, and therefore, significant enhancement of mobility could be obtained. The enhancement of mobility in the present result is more obvious than that in conventional gating using STO (from ~2400 to ~3600 cm$^2$/Vs) and LAO (from ~1000 to ~2000 cm$^2$/Vs) as the dielectric material.[19, 34, 35] The highest mobility is much higher than those in the EDLTs without protection layer (e.g., ~1000 cm$^2$/Vs at IL/STO interface and ~7000 cm$^2$/Vs at IL/KTaO$_3$ interface ),[2-4, 10-13] and that in the EDLT on STO using BN as a protection layer (12000 cm$^2$/Vs).[14]

In order to further demonstrate the enhancement of mobility by IL-assisted field effect, we show the observation of the Shubnikov-de Haas (SdH) oscillation in the magnetoresistance (MR) measurement. Figure 4a shows the variation of resistance $\Delta R =$

$R(B)$ - $R(0)$ at different $V_G$ as a function of magnetic field $B$ oriented perpendicular to the LAO/STO interface, measured at $T$ = 2 K for sample A. For the sample without IL, positive MR and no oscillation is observed for $B$ up to 9 T. After application of IL, $\Delta R$ oscillation is observed for $B$ higher than ~ 4 T even at $V_G$ = 0 V, due to the enhancement of mobility (Fig. 3). $\Delta R$ oscillation is also observed at higher negative $V_G$ for -0.4 ≤ $V_G$ ≤ 0 V. Figure 4b shows numerical derivative of resistance d$R$/d$B$ as a function of the inverse of magnetic field $B^{-1}$. One can see that the oscillations are more visible in d$R$/d$B$ versus $B^{-1}$ plots and the oscillations are periodic in $B^{-1}$ at each $V_G$. With increasing negative $V_G$ and thereby decreasing $n_s$, the significant shift of the main peak of the oscillations is observed.

Figure 4c shows the $R_s$ as function of $B$ for different $T$ at $V_G$ = 0 V. By subtracting a polynomial background for each curve, the amplitude of oscillation is obtained and shown in Fig. 4d. One can see that the amplitude decreases with increasing $T$. The oscillation amplitude $\Delta R$ as a function of $T$ can be described by the relation[19, 20]

$$\Delta R(T) = 4R_0 e^{-\alpha T_D} \alpha T / \sinh(\alpha T) \qquad (1)$$

where $\alpha = 2\pi^2 k_B / \hbar \omega_c$, $\omega_c = eB/m^*$, $k_B$ is the Boltzmann constant, $\hbar$ is the Planck constant, $\omega_c$ is the cyclotron frequency, $e$ is the elementary charge, $B$ is the magnetic field, $m^*$ is the carrier effective mass, $R_0$ is the non-oscillatory component of $R_s$, and $T_D$ is the Dingle temperature. We extracted $\Delta R$ at $B$ = 8.25 T (the peak at $B^{-1}$ = 0.121 T$^{-1}$) and plotted it as a function of $T$ in the inset of Fig. 4d. The fitting of these data by using the equation (1) gives the effective mass $m^*$ = 0.70 ± 0.06 $m_e$, where $m_e$ is free electron mass. The value of $m^*$ observed here is lower than those of previous high-mobility LAO/STO

interfaces,[19-22] and is comparable to that of LAO/STO interface treated by surface control.[22]

## CONCLUSION

In summary, we have performed the IL-assisted electric field effect to tune the transport properties of 2DEG at LAO/STO interface. Depletion of carrier density and the resultant metal-insulator transitions are observed. The LAO capping layer separates the IL from the 2DEG, and therefore, the interface is protected against any electrochemical reaction, which is suggested by facts that the sample surfaces before and after IL gating are unchanged and the metal-insulator transitions are reversible. Therefore, the mobility was significantly enhanced as the carrier density is depleted. Due to the mobility enhancement, SdH oscillations of the conductance were observed. These results suggest that ionic liquid-assisted field effect could be an important avenue to construct high-mobility oxide interfaces which are essential for exploration of quantum phenomena. Moreover, the samples show a significant response on the electric field and exhibit a reversible metal-insulator transition with high ON/OFF ratios, which are potential for device applications such as field effect transistors and electronic switches.

## METHODS

The LAO/STO heterostructures were obtained by depositing LAO thin films on $TiO_2$-terminated (100)-oriented STO substrates using pulsed laser deposition (PLD) system.

The target is single crystalline LAO. Four samples (samples A, B, C and D) with different initial low-$T$ mobilities were synthesized. Sample A was deposited at temperature $T = 760$ °C and has initial mobility of 1110 cm$^2$/Vs at $T = 2$ K. Samples B, C and D were deposited at $T = 650$ °C and have initial mobilities of 2900 cm$^2$/Vs at $T = 3$ K, 3500 cm$^2$/Vs at $T = 2.5$ K and 9300 cm$^2$/Vs at $T = 2.8$ K, respectively. The enhanced initial mobilities for sample B, C and D is attributed to the improvement of crystalline quality of LAO deposited at low $T$. The thickness of samples A, B, C and D are 10, 8, 8 and 8 unit cells (uc), respectively. Sample A was deposited at oxygen partial pressure $P_{O2} = 2\times10^{-3}$ Torr and samples B, C and D were deposited at $P_{O2} = 2\times10^{-4}$ Torr, and then they all were cooled down to room $T$ in the deposition $P_{O2}$ at a rate of 15 °C/min. During the deposition, an *in-situ* reflection high energy electron diffraction (RHEED) was used to monitor the thickness of LAO (Supporting Information Fig. S1). Patterns with a Hall bar geometry and a lateral gate electrode were fabricated to form IL-assisted field-effect devices. In order to fabricate patterned LAO/STO interfaces, before the deposition of LAO, STO substrates were patterned by using conventional photolithography and amorphous AlN films were deposited as a hard mask. Then, the patterned substrates were put into PLD chamber for the deposition of LAO. After deposition, the samples were annealed in a tube furnace at 550 °C for 1 hour in air, in order to remove the oxygen vacancies in STO introduced by high-energy plasma bombardment during the deposition.

After the thin film growth, we did not carry out any processing step such as lithography because exposing the samples to chemicals may deteriorate the quality of surface and interface. The wire connection for transport measurement was done by Al ultrasonic wire bonding. For the connection of gate electrodes, Al wires were first bonded to the lateral gate pads, and then a small amount of silver paint was dropped to cover the gate pad. The silver paint is used to enlarge the area of the gate electrode, enabling the accumulation of ions. A small droplet of an IL, N,N-diethyl-N-methyl-N-(2-methoxyethyl)ammonium bis(trifluoromethyl sulphonyl)imide (DEME-TFSI), was put onto the sample and covered both the conducting channel and the gate electrode. The sample was then placed in the chamber of a Quantum Design Physical Property Measurement System (PPMS) for the transport measurement.

**Conflict of interests**

The authors declare no competing financial interests.

**Acknowledgments**

This work is supported by the National University of Singapore (NUS) Academic Research Fund (AcRF Tier 1 Grant No. R-144-000-346-112 and R-144-000-364-112) and the Singapore National Research Foundation (NRF) under the Competitive Research Programs (CRP Award No. NRF-CRP 8-2011-06 and CRP Award No. NRF-CRP10-2012-02).

**Supporting Information Available**

Characterization of sample growth using reflection high energy electron diffraction (RHEED), electrical characterization of samples B, C and D, and surface characterization of samples before and after gating experiments. This material is available free of charge *via* the Internet at http://pubs.acs.org.

**Figure legends**

**Figure 1.** Optical micrograph, surface and electrical transport characterization of an LAO/STO EDLT. (**a**) Optical micrograph of a Hall-bar pattern on STO using amorphous AlN as the mask and the measurement circuit. (**b**) AFM image for 10 uc LAO grown on patterned STO substrates, measured in the region of Hall bar channel. (**c**) Source-drain current $I_D$ and leakage current $I_G$ as a function of gate voltages $V_G$. (**d**) Resistance as a function of $V_G$ for a device for three scan cycles. Each scan cycle starts from 0 V to 2 V, and then from 2 V to -0.8 V, and finally from -0.8 V to 0 V. The scan speed of $V_G$ is 25 mV/s.

**Figure 2.** Electrical transport characterization of an LAO/STO EDLT. (**a**) The linear-scale sheet resistance versus temperature ($R_s$-$T$) curves at various $V_G$ for sample A. Inset of (**a**) is the logarithmic-scale $R_s$-$T$ curves at low temperature. (**b**) The $n_s$ and (**c**) $\mu_H$ as a function of temperature at various $V_G$ for sample A. The black square is the data without IL on top of the sample.

**Figure 3.** Carrier density $n_s$ and mobility $\mu_H$ at low temperature. (**a**) The $n_s$ and (**b**) $\mu_H$ as a function of $V_G$ for samples A, B, C and D at low temperatures. (**c**) The $\mu_H$ as a function of $n_s$. The data for samples without IL are also shown.

**Figure 4.** Liquid-gated modulation of the Shubnikov-de Haas oscillations for sample A. (**a**) Variation of resistance $\Delta R = R(B) - R(0)$ as a function of magnetic field $B$. (**b**) Numerical derivative $dR/dB$ as a function of the inverse of magnetic field $B^{-1}$. (**c**) Sheet resistance as a function of $B$ for different temperatures at $V_G = 0$ V. (**d**) Oscillatory component of the sheet resistance as a function of $B^{-1}$. Inset of (**d**) is the amplitude of the oscillation at $B = 8.25$ T as a function of temperature. The black squares are the experimental data and the red curve is the fitting line.

**FIGURES**

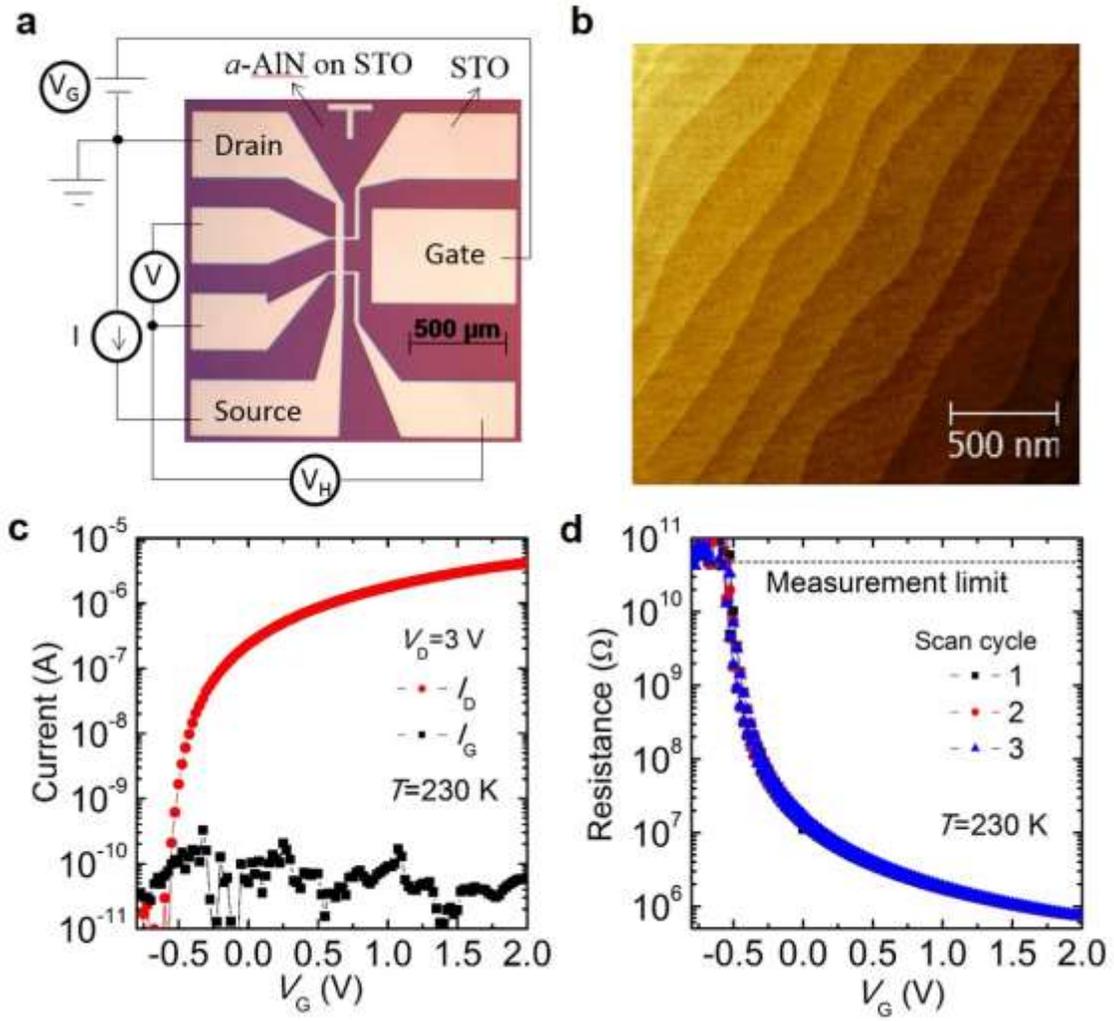

Figure 1

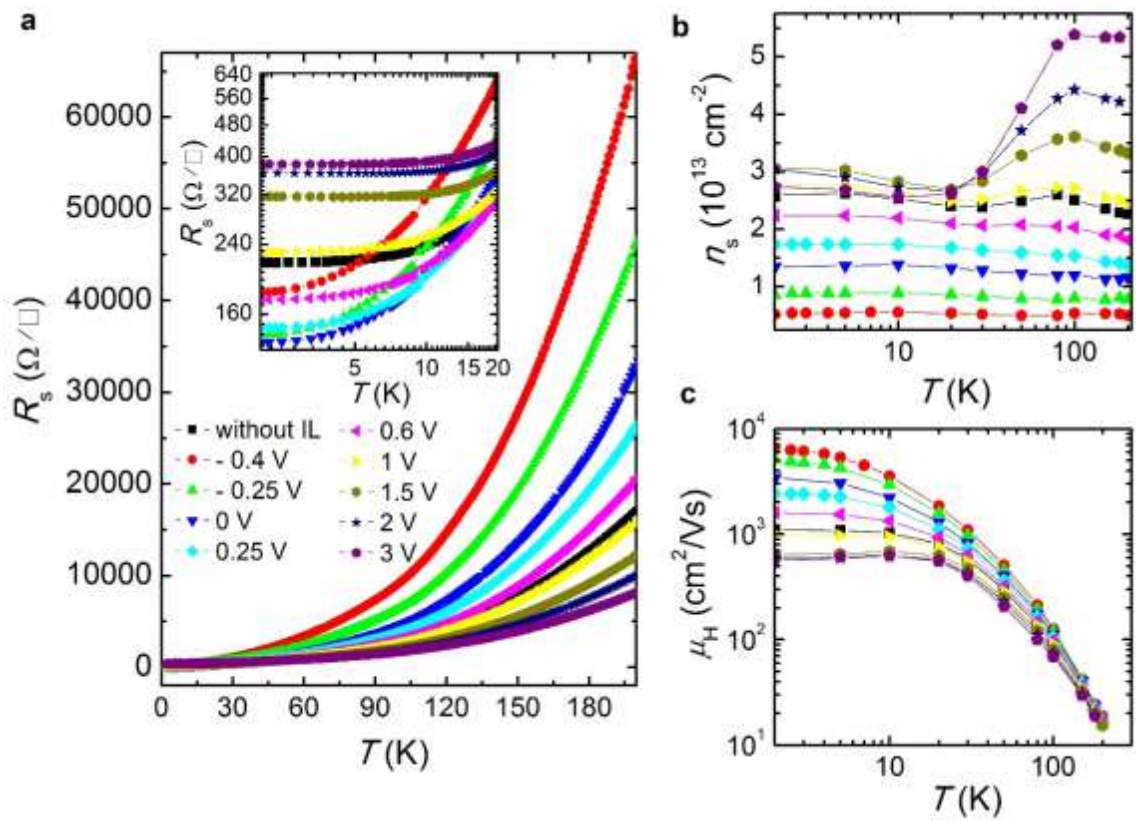

Figure 2

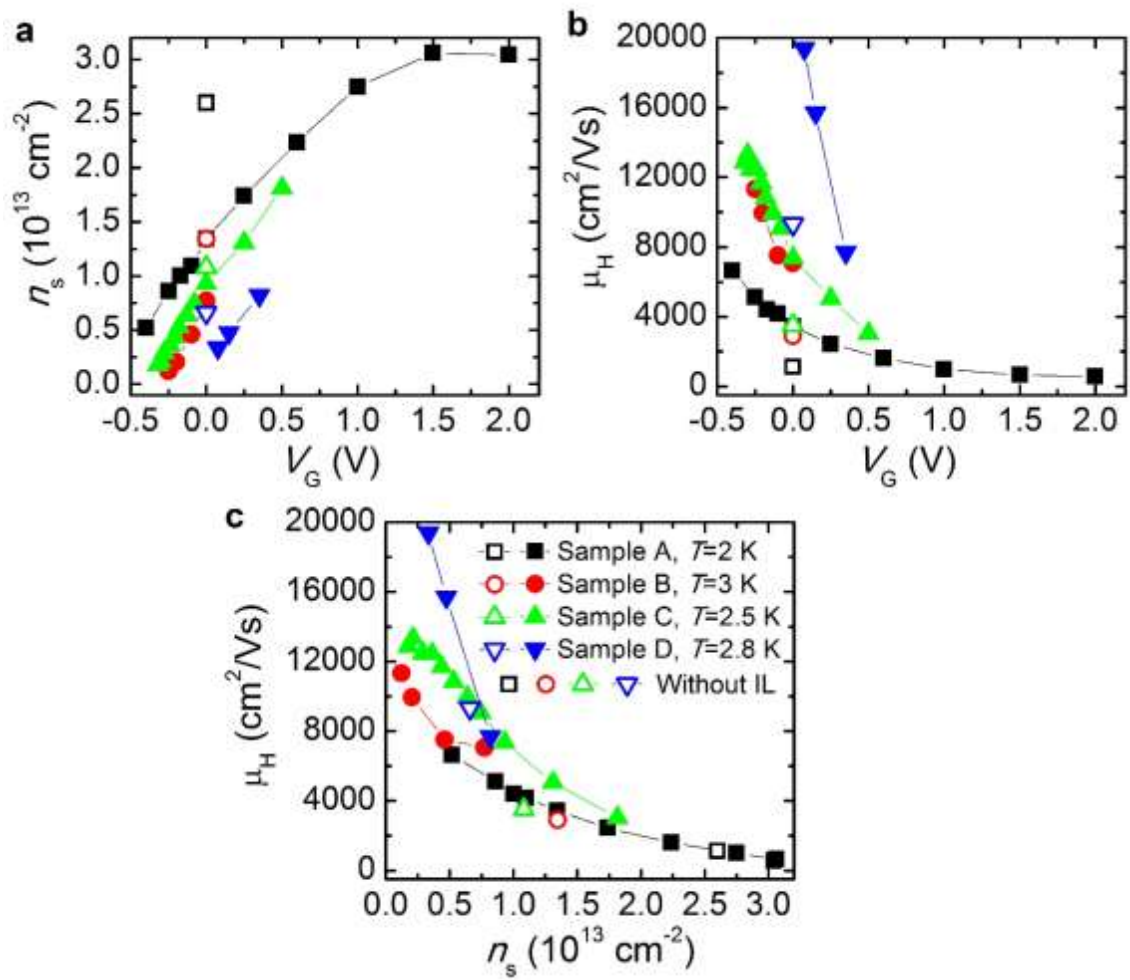

Figure 3

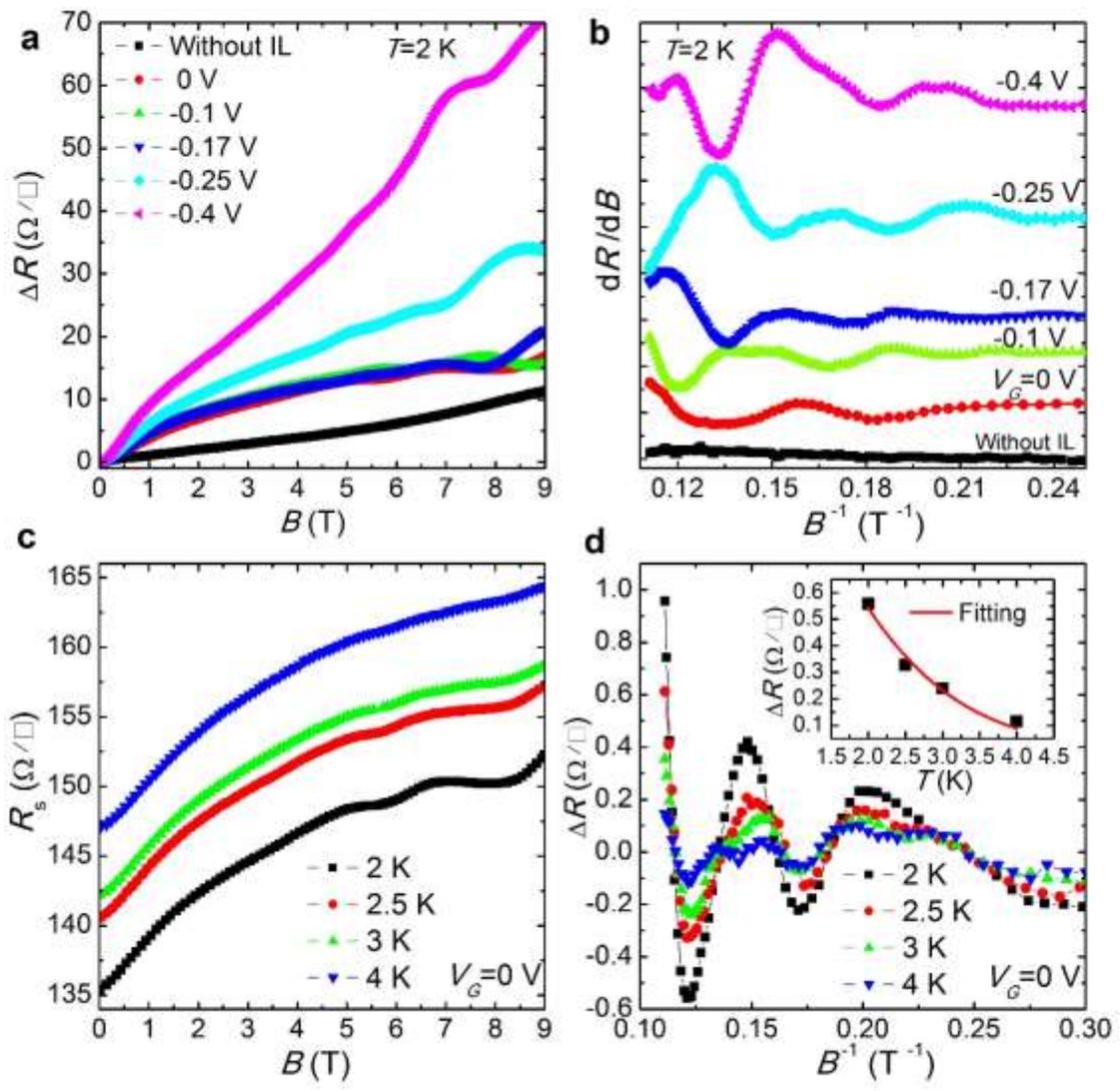

Figure 4